\documentclass[12pt]{article}
\pdfoutput=1
\usepackage{amssymb, amsmath,amsfonts}
\usepackage{multirow}
\usepackage{mathrsfs}
\usepackage{array}
\usepackage{cite}
\usepackage{booktabs}
\usepackage{tikz}
\usepackage{pgfplots}
\usepackage{float}
\usepackage{subfigure, graphicx}
\usepackage[pdftex, bookmarks=true,colorlinks,linkcolor=red,urlcolor=blue,citecolor=blue]{hyperref}
\usepackage{cite}
\usepackage{enumerate}
\usepackage{slashed}
\usepackage{tabularx}
\usepackage{enumerate} 
\usepackage{bm}
\usepackage{tikz}

\makeatletter\@addtoreset{equation}{section}\makeatother

\def\be{\begin{equation}}
\def\ee{\end{equation}}
\def\bea{\begin{eqnarray}}
\def\eea{\end{eqnarray}}
\newcommand{\comment}[1]{{\bf {\textcolor{blue}{ [#1]}}}}

\def\Dslash{\,\,{\raise.15ex\hbox{/}\mkern-12mu D}}
\def\Dbarslash{\,\,{\raise.15ex\hbox{/}\mkern-12mu {\bar D}}}
\def\delslash{\,\,{\raise.15ex\hbox{/}\mkern-9mu \partial}}
\def\delbarslash{\,\,{\raise.15ex\hbox{/}\mkern-9mu {\bar\partial}}}
\def\pslash{\,\,{\raise.15ex\hbox{/}\mkern-9mu p}}
\def\calDslash{\,\,{\raise.15ex\hbox{/}\mkern-12mu {\cal D}}}

\makeatletter\@addtoreset{equation}{section}\makeatother

\renewcommand{\title}[1]{\vbox{\center\LARGE{#1}}\vspace{5mm}}
\renewcommand{\author}[1]{\vbox{\center#1}\vspace{5mm}}
\newcommand{\address}[1]{\vbox{\center\em#1}}

\def\arXiv#1{\href{http://arxiv.org/abs/#1}{arXiv:#1}}
\def\arXiv#1#2{\href{http://arxiv.org/abs/#1}{arXiv:#1}}

\def\doi#1#2{\href{http://doi.org/#1}{#2}}

\begin{document}

\unitlength = .8mm

\begin{titlepage}
\vspace{.5cm}
 
\begin{center}
\hfill 
\hfill \\
\vskip 1cm

\title{Energy transport in holographic \\ non-conformal interfaces}
\vskip 0.5cm
{Yan Liu$^{\,a,b}$}\footnote{Email: {\tt yanliu@buaa.edu.cn}}, {Chuan-Yi Wang$^{\,a, b}$}\footnote{Email: {\tt by2230109@buaa.edu.cn}} and {You-Jie Zeng$^{\,a, b}$}\footnote{Email: {\tt by2330101@buaa.edu.cn}} 

\address{${}^{a}$Center for Gravitational Physics, Department of Space Science,\\ 
Beihang University, Beijing 100191, China}

\address{${}^{b}$Peng Huanwu Collaborative Center for Research and Education, \\Beihang University, Beijing 100191, China}

\end{center}
\vskip 1.5cm

\abstract{We study energy transport in a system of two dimensional conformal field theories exchanging energy across a non-conformal interface involving a localised scalar operator, using  holographic duality. By imposing the sourceless boundary condition, or equivalently, enforcing energy conservation at the interface, we show that the sum of the transmission and reflection coefficients is equal to one. Unlike conformal interfaces, we find that both the energy transmission and reflection coefficients are generally complex and frequency dependent. When the induced metric on the interface brane connects two distinct AdS$_2$ geometries, the transmission coefficient approaches the value expected for a conformal interface in the UV regime at high frequencies and in the IR regime  at low frequencies. In the intermediate frequency range, the transmission coefficient may exhibit oscillatory behavior. Moreover, we present a nontrivial example of a  fully transmissive interface, which exhibits similarities to a topological interface.
}
\vfill

\end{titlepage}






\newpage
\section{Introduction}

Interfaces arise in diverse physical systems, 
including the impurities in samples, quantum wire junctions, 
waveguide systems, surface phenomena in statistical mechanics, and edge modes of topologically ordered systems. Meanwhile, interfaces play crucial roles in the ambient theory, such as Wilson loops probing the phases of gauge theory, $D$-branes in string theory etc. Therefore, characterizing the properties of interfaces is of significant interest across diverse fields. 

When a codimension one interface preserves half of the conformal symmetry, it is termed a conformal interface. In two dimensions, conformal interfaces have attracted substantial research attention. Various  physical quantities, such as interface entropy or the $g$-function \cite{Affleck:1991tk}, effective central charge \cite{Sakai:2008tt},  and energy transport \cite{Quella:2006de, Meineri:2019ycm}, have been extensively explored. 
These quantities are generally distinct but  may be interrelated through simple  inequalities \cite{Karch:2024udk}. This paper focuses on energy transports across the interface. 

Energy transports are key quantities that  characterize the transmission or reflection of energy across an interface. Here, we study a specific case where the two CFTs separated by the interface possess the same central charge, which share similarities as the system of defects or impurities \cite{Oshikawa:1996ww, Lerner2008, Rylands:2016gsf, Billo:2016cpy}. When a pulse of energy is sent from the left side, the transmitted energy to the right is proportional to a  coefficient $\mathcal{T}$, and the reflected portion is proportional to $\mathcal{R}$, with $\mathcal{T}+\mathcal{R}=1$ in the absence of absorption. This process is  governed by the two-point function of the energy-momentum tensor across the interface \cite{Meineri:2019ycm}. 

It has been established in  \cite{Quella:2006de} the 
transmission coefficient  $\mathcal{T}$ and reflection coefficient $\mathcal{R}$ fall  within the interval  $[0,1]$ for conformal interfaces in a class of unitary CFTs, while they break these bounds for nonunitary CFTs. The relation  $\mathcal{T}+\mathcal{R}=1$  always holds. For ICFT, similar results were derived from averaged null-energy condition \cite{Meineri:2019ycm}. 
In strongly coupled systems, 
the AdS/CFT correspondence proves useful \cite{Zaanen:2015oix, Hartnoll:2016apf}. For holographic ICFTs, which are dual to a gravitational system consists of two AdS$_3$ regions with the same cosmological constant separated by an interface brane with tension $\sigma$, the transmission coefficient is given by 
$
 \mathcal{T}
=\left(1+4\pi G L \sigma \right)^{-1}
$ and $\mathcal{R}=1-\mathcal{T}$ \cite{Bachas:2020yxv}. 
These findings were further confirmed in a study of holographic non-equilibrium steady states with an  interface brane  \cite{Bachas:2021tnp}. 
Given that the tension $\sigma$ is constrained to
$[0,1/(4\pi GL)]$ in  AdS/ICFT, new constraints $
\frac{1}{2}\leq\mathcal{T}\leq 1\,,
$ were discovered \cite{Bachas:2020yxv}. 
The permissible range of transmission coefficient can be further tunned from $0$ to $1$ 
 in double brane holographic models  \cite{Baig:2022cnb}   
and Janus configurations \cite{Bachas:2022etu}. Further studies on the energy transports in AdS/ICFT can be found in \cite{Brunner:2015vva, Baig:2023ahz, Biswas:2024xut, Baig:2024hfc, Gutperle:2024yiz}.  

The goal of this paper is to extend beyond conformal interfaces and investigate  energy transport in  scenarios with generic interfaces. We consider a setup where the interface  hosts a localized scalar degree of freedom. In the  holographic context, the bulk system is composed of  the left half AdS$_3$ and the right half AdS$_3$, joined by an interface brane on which a scalar filed is localized. This configuration was previously  
analyzed in \cite{Liu:2024oxg}, where the properties of brane configurations and $g$-functions were systematically investigated, revealing that both the interface entropy and the $g$-functions manifest explicit scale dependence, which directly indicates the symmetry breaking of conformal interface induced by the localized scalar operator. Thus this setup serves as a holographic model for non-conformal interfaces. In this work, we focus on the case of ICFT with identical central charges on both sides of the interface for simplicity, and examine energy transport across the non-conformal interface. 

This paper is organized as follows. In section  \ref{sec2}, we review the construction of AdS/ICFT for a non-conformal interface and analyze the perturbations of the system. In section \ref{sec3}, we first discuss the boundary conditions necessary to obtain the energy transport and then study the  properties of the  transmission coefficients in various examples to identify universal behaviors. Finally, in section \ref{sec:cd} 
we summarize our findings  and discuss open questions. Appendix \ref{app:a} provides useful equations used in the main text, and appendix \ref{app:b}  briefly reproduces the results for the energy transport in conformal interface.

\section{Holographic model of non-conformal interface}
\label{sec2}
In this section, we first review the holographic model for a two-dimensional strongly coupled field theory with a non-conformal interface \cite{Liu:2024oxg},  where a localised scalar operator appears at the interface. We set the left and right CFT of the interface to have the same central charge, resulting in the entire system resembling a holographic model for a defect. Then we analyze the linear fluctuations of the system. 

The gravitational theory is described by the  following action,  
\be
S_\text{bulk}=S_\text{I} +S_\text{II}+S_Q 
\ee
where 
\bea
\begin{split}
S_\text{I}&=\int_{N_\text{I}} d^3x\sqrt{-g_\text{I}}\,\bigg[\frac{1}{16\pi G}\bigg(R_\text{I}+\frac{2}{L_\text{I}^2}\bigg)
    \bigg] , \\
 S_\text{II}   &=\int_{N_\text{II}} d^3x\sqrt{-g_\text{II}}\,\bigg[\frac{1}{16\pi G}\bigg(R_\text{II}+\frac{2}{L_\text{II}^2}\bigg)
    \bigg] ,\\
S_{Q}&=\frac{1}{8\pi G}\int_Q d^2y\sqrt{-h}\, \bigg[\big(K_\text{I}-K_\text{II}\big)-(\partial\phi)^2-V(\phi)\bigg].
\end{split}
\eea
The coordinates on $N_\text{I}$ ($N_\text{II}$) are $x_\text{I}^a=(u_\text{I}, t_\text{I}, x_\text{I})$ ($x_\text{II}^a=(u_\text{II}, t_\text{II}, x_\text{II})$). We choose the AdS boundary of $N_\text{I}$ to lie in the region $x_\text{I}<0$ while boundary of $N_\text{II}$ is in the region $x_\text{II}>0$. The boundary of interface brane $Q$ is located at  $x_\text{I}=x_\text{II}=0$. Assuming the coordinates on $Q$ are $y^\mu=(t, z)$, the profile of the brane $Q$ in the bulk is parameterized by  $x_\text{I}^a=x_\text{I}^a(y^\mu)$ and $x_\text{II}^a=x_\text{II}^a(y^\mu)$. The continuous condition of the metric on $Q$, as well as the equation of motion on $Q$,  constrain the permissible embeddings. The equations of motion for the system can be found in appendix \ref{app:a1}. 

Here, we assume that CFT$_\text{I}$ and CFT$_\text{II}$ have identical central charge, i.e. $L_\text{I}=L_\text{II}=L$. The interface in the field theory can be thought as a defect or impurity within the system. We focus on the zero-temperature system.  The bulk geometry for $N_\text{I}$ and $N_\text{II}$ is the planar AdS$_3$ solution
\begin{align}
ds^2_A=\frac{L^2}{u_A^2}\bigg[-dt_A^2+
dx^2_A+du_A^2
\bigg]\,, ~~~ A=\text{I\,,~~II}\,.
\end{align}
We have the equation for $Q$ as
 \be
 x_\text{I}=\psi(z)\,,~~~x_\text{II}=-\psi(z)
 \ee
 and the equations of motion for the scalar field or equivalently the null energy condition constraints $\psi''(z)\leq 0$.  
 
For the equilibrium configuration, we have the solution with  $t_\text{I}=t_\text{II}=t, ~u_\text{I}=u_\text{II}=z$. 
 The resulting equations are
  \be \label{eq:bgeqphiV}
 \phi'^2=-\frac{L\, \psi''}{2z\sqrt{1+\psi'^2}}\,,~~~
 V(\phi)=\frac{2\psi'(1+\psi'^2)-z\psi''}{2L(1+\psi'^2)^{3/2}}\,.
 \ee 
These equations are the same as the configuration in BCFT setup, and the ICFT could be understood as an unfold of BCFT at the leading order \cite{Kanda:2023zse, Liu:2024oxg}. In the following, we consider the linear fluctuations of the entire system and derive the energy transport from the fluctuations. 

\subsection{Solution at first order and energy transport}
\label{sec:sts}

To compute the holographic transport coefficient, we consider the first order perturbation of the equation of motion. The first order perturbation of the bulk metric on the left and right halves of the bulk is given
by
\be
ds^2=ds^2_A+[ds^2]^{(2)}_A\,,~~~~~~ A=\text{I\,,~~II}\,
\ee
where  
\begin{align}
\label{eq:pertur}
\begin{split}
&[ds^2]^{(2)}_\text{I}=4GL\epsilon
\left[
e^{i\omega (t_\text{I}-x_\text{I})}(dt_\text{I}-dx_\text{I})^2+
\mathcal{R}e^{i\omega (t_\text{I}+x_\text{I})}(dt_\text{I}+dx_\text{I})^2
\right] +c.c. \,  , \\
&[ds^2]^{(2)}_\text{II}=4GL\epsilon
\mathcal{T}e^{i\omega (t_\text{II}-x_\text{II})}(dt_\text{II}-dx_\text{II})^2 +c.c. \,  .
\end{split}
\end{align}
Here, $\mathcal{R},\mathcal{T}$ are 
the relative amplitudes of the reflected and transmitted waves, respectively, and are referred to as the reflection and  transmission coefficients for convenience. Their precise physical meaning will be explained in detail in the following. The term $c.c.$ denotes complex conjugation. 

From the dictionary of AdS/CFT, the dual energy momentum tensor is given by 
\be
T_{\alpha\beta} =\frac{1}{4GL} g_{\alpha\beta}^{(2)}\,.
\ee
Thus, the injected, reflected, and transmitted energy fluxes are
\be\label{eq:tretr}
 T_\text{in} /\epsilon= e^{i\omega (t_\text{I}-x_\text{I})}+c.c.\,,~~
 T_\text{re}/\epsilon= \mathcal{R}e^{i\omega (t_\text{I}+x_\text{I})}+c.c.\,,~~
 T_\text{tr} /\epsilon= \mathcal{T}e^{i\omega (t_\text{II}-x_\text{II})}+c.c.\,.
\ee
At the location of the interface  $x_\text{I}=x_\text{II}=0$, 
the conservation of energy 
$
 T_\text{in}  =
 T_\text{re}  +
 T_\text{tr} 
$
can be rewritten as
$
(\mathcal{R}+\mathcal{T}-1) e^{i\omega t_\text{I}}+
c.c.=0  \, .
$
This relation implies  
\be
\label{eq:rt=1}
\mathcal{R}+\mathcal{T}=1\,.
\ee
In general, $\mathcal{R}, \mathcal{T}$ are complex and depend on $\omega$. Among the four real components of $\mathcal{R}, \mathcal{T}$, only two are independent due to the constraint in Eq. \eqref{eq:rt=1}. 

The coefficients can be expressed in the form of 
\be\label{eq:phit}
\mathcal{R}=|\mathcal{R}|e^{i\phi_r}\,,~~~ \mathcal{T}=|\mathcal{T}|e^{i\phi_t}\,.\ee From \eqref{eq:tretr}, it is evident that the phases differ among the injected, reflected and transmitted waves, analogous to the behavior of electromagnetic waves at the interface of different media.  
Since $\mathcal{R},\mathcal{T}$ are generally  complex, the reflection and transmission ratios are defined as,
\begin{align}
\label{eq:retrco}
\begin{split}
\tilde{\mathcal{R}}&=\frac{\text{reflected energy flux}}{\text{injected energy flux}}=
\frac{ T_\text{re}  }{ T_\text{in} }=
\frac{
|\mathcal{R}|\cos (\omega t+\phi_r)
}{
\cos (\omega t)
}  \,  , \\
\tilde{\mathcal{T}}&=\frac{\text{transmitted energy flux}}{\text{injected energy flux}}=
\frac{ T_\text{tr}  }{ T_\text{in}  }=
\frac{
|\mathcal{T}|\cos (\omega t+\phi_t)
}{
\cos (\omega t)
} \, .
\end{split}
\end{align}
These ratios are both frequency- and time-dependent. From \eqref{eq:rt=1}, we have
\be
\label{eq:conservation}
\tilde{\mathcal{R}}+\tilde{\mathcal{T}}=1  \, .
\ee
Several observations can be made:
\begin{itemize}
\item When $\mathcal{R},\mathcal{T}$ are real (i.e. $\phi_r=\phi_t=0$), we have 
\be
\tilde{\mathcal{R}}=\mathcal{R}\,,~~~\tilde{\mathcal{T}}=\mathcal{T}\,.
\ee
They are interpreted as the reflection and transmission coefficients for conformal interface \cite{Bachas:2020yxv}. However, 
in general, they are complex, and a phase shift occurs in the energy fluxes. 
\item 
The instantaneous values of 
$\tilde{\mathcal{R}},\tilde{\mathcal{T}}$ depend on the phase $\omega t$ of the injected wave at the interface.  
When $\omega t=(n+\frac{1}{2})\pi$ for integer $n$, 
at the interface 
the injected wave $\langle T_\text{in} \rangle=2\epsilon\cos (\omega t)=0$, and  $\langle T_\text{re} \rangle=\langle T_\text{tr} \rangle$. 
\item 
It is convenient to decompose the coefficients as \be \mathcal{R}=\mathcal{R}'+i\mathcal{R}''\,,~~~~ \mathcal{T}=\mathcal{T}'+i\mathcal{T}''\ee
where $\mathcal{R}',\mathcal{R}'',\mathcal{T}',\mathcal{T}''$ are real and satisfy
\be
\mathcal{R}'+\mathcal{T}'=1\,,~~~~\mathcal{R}''+\mathcal{T}''=0\,. 
\ee
 From \eqref{eq:retrco}, 
the reflection and transmission ratios become
 \be
\tilde{\mathcal{R}}=\mathcal{R}'-\mathcal{R}''\tan(\omega t)\,,~~~\tilde{\mathcal{T}}=\mathcal{T}'-\mathcal{T}''\tan(\omega t)\,.~~~
 \ee
We define the {\em averaged} transmission and reflection coefficients by performing the principle integration from $-\frac{\pi}{2\omega}$ to $\frac{\pi}{2\omega}$,
\bea
\label{eq:aveT}
\begin{split}
\langle\tilde{\mathcal{R}}\rangle =\mathcal{R}'\,,~~~~
\langle\tilde{\mathcal{T}}\rangle =\mathcal{T}'\,.~~~~
\end{split}
\eea
In the low frequency limit, the integration region becomes sufficiently long such that the measured quantities align with the average values. 
\end{itemize}

\subsubsection{The equations of perturbations}
\label{sec:eqpertur}

In this subsection, we show a detailed procedure for calculating the coefficients \eqref{eq:pertur} using holography. We focus on the fluctuation of the interface brane that glues the left and right bulk together. By solving the resulting equations with appropriate boundary conditions, we derive  the transport coefficients. 

We adopt the following coordinate on $Q$ 
\be
t=\frac{1}{2}(t_\text{I}+t_\text{II})\,,~~~
z=\frac{1}{2}(u_\text{I}+u_\text{II})\,.
\ee
These gauge conditions are the same as those were used in \cite{Banerjee:2024sqq}.\footnote{Note that the setup is different from AdS/DCFT in \cite{Erdmenger:2014xya, Erdmenger:2015xpq} where the condition $t_\text{I}=t_\text{II}$ and $u_\text{I}=u_\text{II}$ are used, even in the presence of fluctuations. While the latter approach yields four independent fluctuations for the string worldsheet, it does not lead to a consistent set of  fluctuations \cite{Bachas:2020yxv}.}  The above gauge fixings reduce two of the six worldsheet diffeomorphisms of the interface brane. The perturbed brane is then  parameterized as 
\begin{align}
\begin{split}
&t_\text{I}=t+\frac{G\epsilon}{L}e^{i\omega t}\lambda(z)+c.c.\, ,\ 
x_\text{I}=\textcolor{black}{\psi}(z)+
\frac{2G\epsilon}{L}e^{i\omega t}h_\text{I}(z)+c.c.\, ,\ 
u_\text{I}=z+\frac{2G\epsilon}{L}e^{i\omega  t}\xi(z)+c.c.\,  ,\\
&t_\text{II}=t-
\frac{G\epsilon}{L}e^{i\omega t}\lambda(z)+c.c.\, ,\ 
x_\text{II}=-\textcolor{black}{\psi}(z) +
\frac{2G\epsilon}{L}e^{i\omega t}h_\text{II}(z)+c.c.\, ,\ 
u_\text{II}= z-\frac{2G\epsilon}{L}e^{i\omega t}\xi(z)+c.c.\,   ,
\end{split}
\end{align}
and the perturbed scalar is given by
\begin{align}
\phi(z)=\phi_{bg}(z)+4G\epsilon\, e^{i\omega t} p(z)+c.c.
\end{align} 
where $t$ and $z$ are the intrinsic coordinates of the brane, and $\phi_{bg}(z)$ is the background scalar field. Note that the above perturbations break spatial inversion symmetry.  

The system has five independent perturbed fields, i.e. $\lambda(z), h_\text{I}(z),h_\text{II}(z), \xi(z),$ $p(z)$.  
Substituting these fluctuations into the  system's equations yields seven equations, of  which only five are independent. We have verified that the remaining two equations are consistently satisfied by these five equations. Specifically, the perturbed equations of motion can be simplified into  two independent sourced ODE for $\xi(z)$ and $p(z)$, along with three coupled ODEs where 
$\lambda(z), h_\text{I}(z),h_\text{II}(z)$ are determined by $\xi(z)$ and $p(z)$. We have 
\bea
\xi''-\left( \frac{2}{z}+\frac{\psi''}{\psi'}  \right)\xi'+
\left[
\frac{2}{z^2}+(1+\psi'^2)\omega^2+
\frac{
(1-\omega^2z^2)\psi''
}{z\psi'}
\right]\xi &=& 
\frac{\left(z\psi''-\psi'\right)z}{\psi'}
\left(\bm{I}+\bm{R}-\bm{T}
\right)\,, ~~~~~ ~~  \label{eq:xiODE}
\\
p''-\frac{\psi'\psi''}{\psi'^2+1}p' 
+c_1p &=& ic_2(\bm{R}-\bm{T})+ic_3\bm{I} \,,~~~~~~  \label{eq:psiODE}
\eea
where 
\begin{align}
\bm{I}=e^{-i\omega \psi}  \,, \ \ ~~~~
\bm{R}=\mathcal{R}\,e^{+i\omega \psi}  \,, \ \ ~~~~
\bm{T}=\mathcal{T}\,e^{+i\omega \psi}  \,, \end{align}
and the coefficients $c_1,c_2, c_3$
  in \eqref{eq:psiODE} depend on the brane profile 
$\psi(z)$ and its derivatives (see appendix \ref{app:eom} for explicit forms). 
The other three fields are determined by 
\begin{align}
\lambda(z)&=
\lambda\, [\xi(z),\bm{I},\bm{R},\bm{T}] \label{eq:lambdaODE} \,  ,\\
h_\text{I}(z)&= h_\text{I}\, [\xi(z),\xi'(z),p(z),p'(z),\bm{I},\bm{R},\bm{T}] \label{eq:h1ODE} \,  ,\\
h_\text{II}(z)&= h_\text{II}\, [\xi(z),\xi'(z),h_\text{I}(z),\bm{I},\bm{R},\bm{T}] \label{eq:h2ODE}  \, .
\end{align}
The explicit form of these three equations are provided in appendix \ref{app:eom}. 

The solution for $\xi(z)$ in Eq. \eqref{eq:xiODE} takes the  form\footnote{\label{footnote:sourceless}An  exception occurs when the brane profile takes the form $\psi=-a z^2$, where the source term in Eq. \eqref{eq:xiODE} vanishes. This case  will be discussed in Sec. \ref{subsec:IRflat}. 
} 
\be
\label{eq:gensolxi}
\xi(z)=\xi_l(z)+\xi_s(z)\,,~~~~~~~
\xi_s(z)=-\frac{z}{\omega^2}
\left(
\bm{I}+\bm{R}-\bm{T}\right)
\ee
where $\xi_l$ is the generic solution of the homogeneous part of the ODE \eqref{eq:xiODE},  and $\xi_s(z)$ is a particular solution arising from the source term for a generic brane profile $\psi(z)$. 

For a given background brane profile, $\xi(z), p(z)$ can be obtained by solving Eqs. \eqref{eq:xiODE} and  \eqref{eq:psiODE} 
with appropriate boundary conditions: the infalling boundary conditions for the  homogeneous solutions and the sourceless boundary conditions near the AdS  boundary. Substituting these solutions into Eqs. \eqref{eq:lambdaODE}, \eqref{eq:h1ODE} and \eqref{eq:h2ODE} yields the solutions for remaining three fields under the same boundary conditions. These boundary conditions will constrain the system for a specific allowed values of $\mathcal{R}$ and $\mathcal{T}$.

\section{Energy transports for different brane profiles}
\label{sec3}

In this subsection, we will study the energy transport coefficients for various different brane profiles, following an explanation of  the systematic approach to solving the system. In the first set of examples, consisting of one numerical and two analytical cases, the induced metric on the interface brane $Q$ is asymptotically AdS$_2$ in both the UV and IR regions. We identify universal  features in the energy transport coefficients across these scenarios. In the second case, where the induced metric on $Q$ is asymptotically AdS$_2$ in the UV while asymptotically flat in the IR, the system shows behavior analogous to a  topological interface, exhibiting fully  transmissive energy flux. 

\subsection{On the UV boundary condition for  asymptotic AdS$_2$ solution}
\label{subsec:bc}

In this subsection, we focus on solutions where the induced metric on the interface brane $Q$ is asymptotically AdS$_2$. 
A class of such solution corresponds to the brane profile $\psi(z)$ with a UV expansion as $z\to 0$, 
\be\label{eq:pzAB}
\psi(z)=Az+Bz^2+...  \, ,
\ee 
where $A\not=0$ and $B\leq 0$. 
For a given profile of $Q$, the scalar field $\phi(z)$ and the potential $V(\phi)$ are  determined from Eq. \eqref{eq:bgeqphiV}.  
The localized scalar operator in the dual ICFT has a conformal dimension of $1/2$ for $B<0$ and $1$ for $B=0$  while the coefficient of $\psi(z)$  in front of $z^3$ nonzero. 
Below, we show that under the sourceless boundary condition for the fluctuations, the relation $\mathcal{R}+\mathcal{T}=1$ holds, consistent with energy conservation as  discussed in Sec. \ref{sec:sts}.

With the above asymptotic behavior, 
the solution to Eq. \eqref{eq:xiODE} has a UV expansion
\be\label{eq:xibnd}
\xi(z)=\xi_1\, z+\xi_2\, z^2+
\frac{1}{2}\left[
-1-\mathcal{R}+\mathcal{T}-\omega^2 (1+A^2)\xi_1
+\frac{2B\xi_2}{A}
\right]z^3+\cdots\,,
\ee
\textcolor{black}{where $\xi_1$ and $\xi_2$ are integration constants, 
formed by combining  the homogeneous solution's integration constants with the asymptotic expansion terms of the particular solution in \eqref{eq:gensolxi}.}

For $B<0$, the solution to Eq. \eqref{eq:psiODE} has the following UV expansion, 
\begin{align}
p(z)=p_1\sqrt{z}\log(z)\left(1+\mathcal{O}(z)\right)
+p_2\sqrt{z}\left(1+\mathcal{O}(z)\right)
   \,  , \label{eq:p(z)whenb<0}
\end{align}
where $p_1$ and $p_2$ are two integral constants. 
Here, $p(z)$ in \eqref{eq:p(z)whenb<0} represents the asymptotic expansion of the complete solution. Unlike
our previous analysis of $\xi(z)$ above, the contribution from the sourced term of \eqref{eq:psiODE} to $p(z)$ only appears in higher-order terms of the expansion.
For $B=0$ while the coefficient in front of $z^3$ nonzero, the UV expansion is
\begin{align}
p(z)=p_1+p_2 z+\cdots
   \,  ,
\end{align} 
where $p_1$ and $p_2$ are again integration constants. For $B<0$, applying the sourceless condition $p_1=0$ yields, $
p(z)=p_2\sqrt{z}\left(1+\mathcal{O}(z)\right).
$ 
For $B=0$ while the coefficient in front of $z^3$ nonzero, the sourceless condition gives $p(z)=p_2\, z+\cdots$. Note that in \eqref{eq:pzAB} if the nonzero higher order term of $z^n$ first starts with an integer $n$ of $n\geq 2$, the scalar fluctuation have higher conformal dimension $(n-1)/2$. 

As noted  in \cite{Bachas:2020yxv}, the boundary values of the functions $\xi, \lambda, h_\text{I}, h_\text{II}$ as $z\to 0$ correspond to the sources of operators in the dual ICFT. Using the UV expansions of $\psi(z),\xi(z), p(z)$, along with the sourceless conditions of other functions, we can derive $\mathcal{R}+\mathcal{T}=1$. 
First, substituting $\psi(z),\xi(z)$ into Eq.  \eqref{eq:lambdaODE}, the leading term for $\lambda$ is 
\be
\lambda(0)= -\frac{2i\xi_1}{\omega }  \, .
\ee
The condition $
\lambda(0)=0\, 
$
requires $\xi_1=0$. 
Second, substituting $\psi(z),\xi(z)$ into Eq. \eqref{eq:h2ODE}, we find 
\be
h_\text{II}(0)=-h_\text{I}(0)-
\frac{2\xi_2}{A\omega^2}  \, .
\ee
The conditions $
h_\text{I}(0)=h_\text{II}(0)=0  \, 
$
require $\xi_2=0$. Third, substituting  $\psi(z),\xi(z), p(z)$ into Eq.  \eqref{eq:h1ODE}, we find
\be
h_\text{I}(0)=\frac{i}{\omega^3}
(\mathcal{R}+\mathcal{T}-1)
\, .
\ee
The condition $
h_\text{I}(0)=0  \, 
$
requires 
\be\label{eq:R+T=1}
\mathcal{R}+\mathcal{T}=1  \,  .
\ee
Note that in Sec. \ref{sec:sts}, we have shown that this relation is a consequence of energy conservation across the interface. 
Here, we have demonstrated that in  holography, the relation  $\mathcal{R}+\mathcal{T}=1$ is enforced by the sourceless boundary condition near the UV AdS$_2$ boundary.

The sourceless boundary conditions constrain  the expansion coefficients of $\xi$ in Eq.  \eqref{eq:xibnd} to vanish as $z\to 0$, i.e. \be
\label{eq:sourcelessbnd}
\xi_1=\xi_2=0\,.\ee 
For the profile with asymptotic behavior given in Eq. \eqref{eq:pzAB}, using the condition \eqref{eq:R+T=1}, the inhomogeneous solution $\xi_s$ in Eq. \eqref{eq:gensolxi} 
behaves as $z\to 0$ 
\be
\xi_s=-\frac{2(1-\mathcal{T})}{\omega^2}z+\frac{2iA\mathcal{T}}{\omega}z^2+\cdots\,.
\ee
From the sourceless condition \eqref{eq:sourcelessbnd}, 
the homogeneous solution $\xi_l$ must satify 
\be
\label{eq:asyxil}
\xi_l=\frac{2(1-\mathcal{T})}{\omega^2}z-\frac{2iA\mathcal{T}}{\omega}z^2+\cdots\, 
\ee
when $z\to 0$\,.
By imposing the infalling boundary condition for $\xi_l$ near the IR and the above UV boundary condition, we can solve the system and obtain the final solution for $\xi_l$ and the transmission coefficient $\mathcal{T}$. 

\subsection{An example from numerical studies
}
\label{subsec:sol1}

In this subsection, we consider the following background profile of the interface brane 
\begin{align}
\label{eq:profile1}
\psi(z) =az+\frac{(b-a)z^2}{1+z}  \,  ,
\end{align} 
where the brane geometry are asymptotically AdS$_2$ in both the UV and IR limits. The null energy condition on the brane constrains $a\geq b$. 
The parameters $a$ and $b$ are related to the UV and IR limits of the interface entropy by 
\begin{align}
a=\sinh{\frac{3S^\text{UV}_\text{iE}}{c}} \, ,~~~~
b=\sinh{\frac{3S^\text{IR}_\text{iE}}{c}} \, ,
\end{align}
where the interface entropy is defined in \cite{Liu:2024oxg}, and $c=3L/2G$ is the central charge of the left/right CFT. We  restrict our discussion to the region $b>0$ to ensure a positive interface entropy. 
The UV expansion of this profile matches exactly with Eq. \eqref{eq:pzAB} upon noticing that $A=a$ and $B=b-a$.

On the interface brane, we have a nontrival scalar field. The boundary values give  information about the scalar operator located at the interface. The source of the scalar operator $\langle\mathcal{O}\rangle$ vanishes, and its VEV is 
$
\langle\mathcal{O}\rangle=
2\sqrt{2(a-b)L}
/
(1+a^2)^{1/4}
  \,  .
$ 
We will study the dependence of transport coefficients on three free dimensionless parameters: $a,b$, and  $\omega/\langle\mathcal{O}\rangle^2$.

We begin by solving the homogeneous part of the ODE given in Eq. \eqref{eq:xiODE}. The solution $\xi_l$ is obtained by imposing the infalling boundary condition near the horizon. In the IR limit $z\to\infty$, the asymptotic behavior of $\xi_l$  is 
\begin{align}
\label{eq:bndxil}
\xi_l= \xi_0\, e^{-i\omega\sqrt{1+b^2}z}\left(
z+\frac{
i(a-b)\omega\sqrt{1+b^2}
}{
b
}
+\mathcal{O}(z^{-1})
\right)  \,  ,
\end{align}
where $\xi_0$ is an integral constant, and we have dropped the out-going wave term  $e^{+i\omega\sqrt{1+b^2}z}$. The homogeneous part of the ODE  \eqref{eq:xiODE} can be solve using the boundary condition specified in Eq. \eqref{eq:bndxil}. By matching the asymptotic behavior of $\xi_l$ with Eq.  \eqref{eq:asyxil}, we can determine the transmission coefficient $\mathcal{T}$. 
Additionally, the scalar fluctuation can be obtained by imposing the infalling boundary condition discussed in appendix \ref{app:a3} and the sourceless boundary condition. This approach allows us to determine all the fluctuations in the system. 

When $b=a$, the profile simplifies to $\psi(z)=az$, corresponding to a conformal interface previously studied in \cite{Bachas:2020yxv}. In this case, the solution to Eq. \eqref{eq:xiODE} is   
\be
\xi_l=\xi_0 z e^{-i\omega z\sqrt{1+a^2}}
\,.
\ee
Matching this solution with Eq.  \eqref{eq:asyxil} in the limit $z\to 0$, we obtain 
\begin{align}
\xi_0=\frac{2a(\sqrt{1+a^2}-a)}{\omega^2} \,, ~~~~\mathcal{T}=1+a^2-a\sqrt{1+a^2} \,.
\end{align}
This result agrees exactly with the findings in \cite{Bachas:2020yxv}, as can be verified by setting \eqref{BachasT} with $\tan\alpha=\tan\beta=a$, as calculated in appendix \ref{app:b}.

\begin{figure}[h!]
\begin{center}
\includegraphics[width=0.46\textwidth]{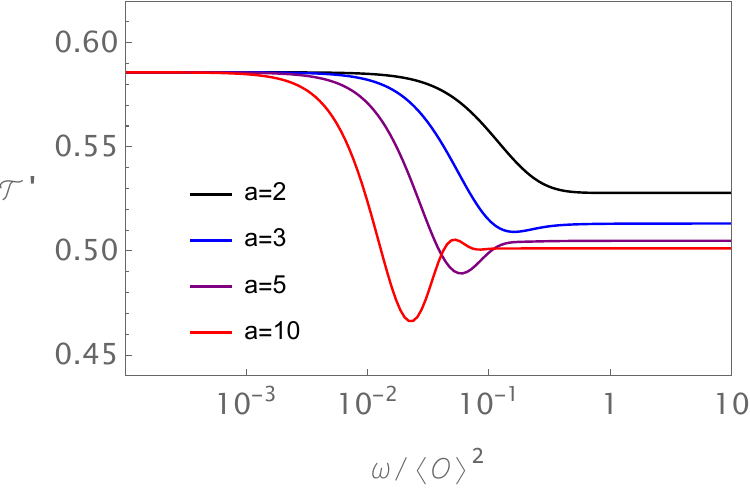}
~~~~~
\includegraphics[width=0.46\textwidth]{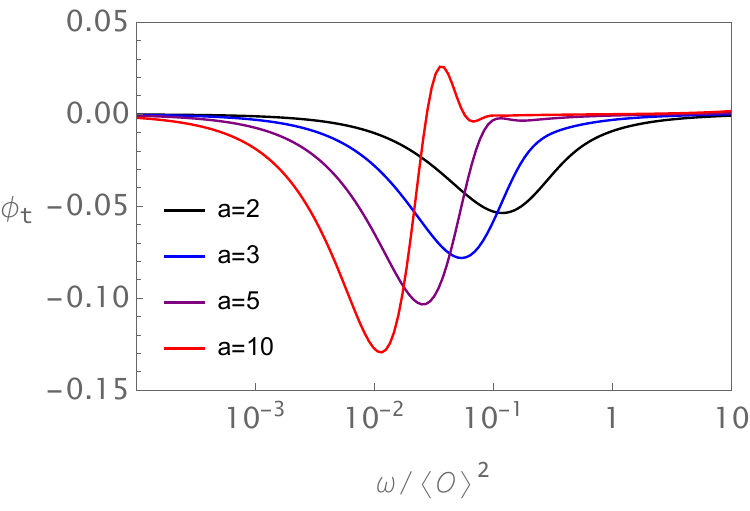}
\\
\includegraphics[width=0.46\textwidth]{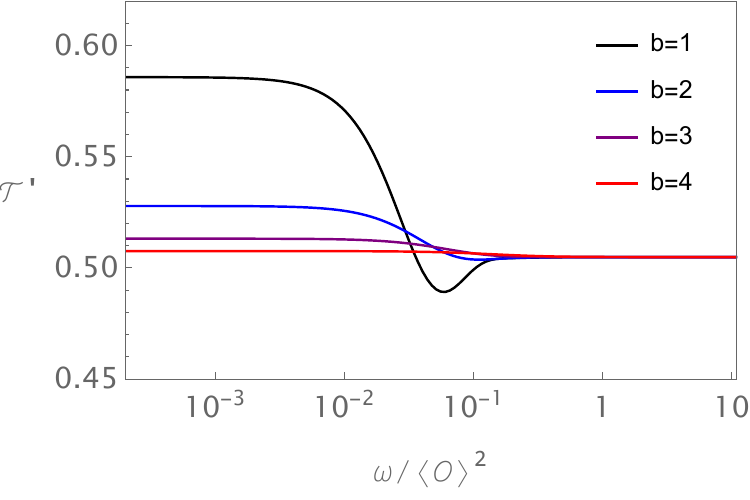}
~~~~~
\includegraphics[width=0.46\textwidth]{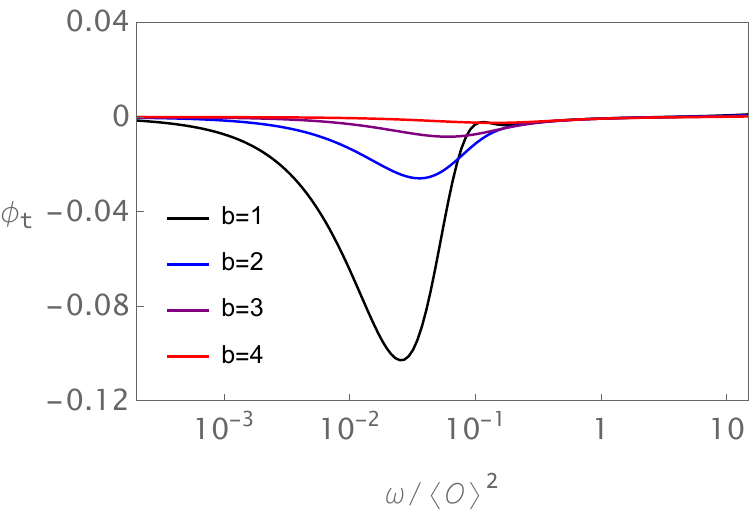}
\end{center}
\vspace{-0.3cm}
\caption{\small  The plots of $\mathcal{T}'$ and $\phi_t$ 
as functions of $\omega/\langle \mathcal{O}\rangle^2$
for varying $a$ with fixed $b=1$ ({\em upper two plots}), and varying $b$ with fixed $a=5$ ({\em lower two plots}).
}
\label{fig:Toab}
\end{figure}

In the following we consider the case $a>b$ and analyze the behaviors of the averaged transmission coefficient $\mathcal{T}'$ defined in \eqref{eq:aveT} and the phase shift of the transmission wave $\phi_t$ defined in  \eqref{eq:phit}.  Fig. \ref{fig:Toab} illustrates the dependence of $\mathcal{T}'$ and $\phi_t$ on  $\omega/\mathcal{O}^2$ for different values of $a$ with $b=1$ in the upper two plots and for different values of $b$ with fixing $a=5$ in the lower two plots. 
From Fig. \ref{fig:Toab}, we observe several key features of the transport coefficients. 
Unlike the conformal interface studied in \eqref{BachasT}, the transmission coefficient in this case is generally complex, and the averaged transmission coefficient  $\mathcal{T}'$ exhibits a dependence on 
the frequency $\omega$ of the incident wave. 
Furthermore, $\mathcal{T}'$ and the phase $\phi_t$ approach constant values in both the low- and high-frequency limits of  $\omega/\mathcal{O}^2$.
In the low-frequency regime, the averaged transmission coefficient $\mathcal{T}'$ converges to a constant that depends solely  on the parameter $b$. This value coincides with the transmission coefficient for the brane profile $\psi(z)=b z$ of conformal interface. Simultaneously, the phase $\phi_t$ approaches zero, indicating that the transmission coefficient $\mathcal{T}$ is entirely governed by the IR region of the interface brane. Conversely, in the high-frequency regime, $\mathcal{T}'$ approaches to a constant determined by the parameter $a$, matching the transmission coefficient for the brane profile $\psi(z)=a z$ of the conformal interface. The phase $\phi_t$ also tends to zero in this limit,  signifying that $\mathcal{T}$ is dominated by the UV region of the interface brane.

The UV and IR behaviour of the brane can be approximated by a straight line. Thus, the UV and IR limits of $\mathcal{T}'$ always satisfy the bound proposed in  \cite{Bachas:2020yxv}.
The null energy condition requires $a\ge b$ ensuring that the UV value of $\mathcal{T}'$ is never larger than its IR value. 
In the intermediate frequency region, $\mathcal{T}'$ and $\phi_t$ may exhibit  oscillatory behavior. As $a$ increases or $b$ decreases, the number of oscillations  increases. 
Notably, in this intermediate frequency range, the averaged transmission coefficient $\mathcal{T}'$ violates the lower bound proposed in \cite{Bachas:2020yxv}. 

\begin{figure}[h!]
\begin{center}
\includegraphics[width=0.46\textwidth]{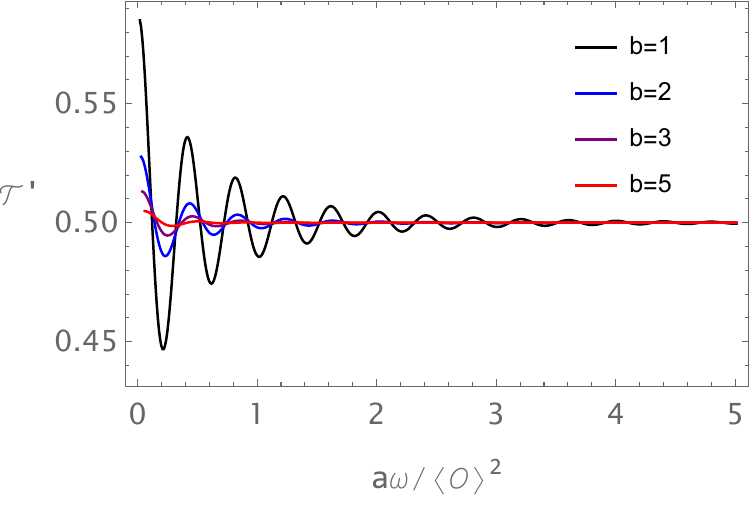}
~~~~~
\includegraphics[width=0.46\textwidth]{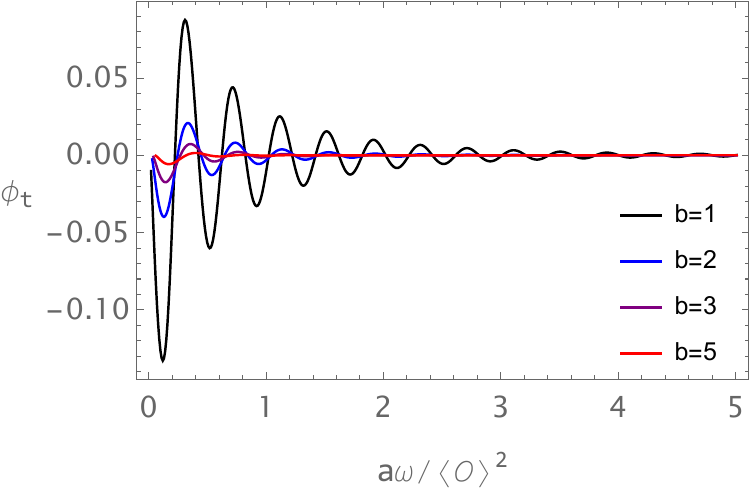}
\\
\includegraphics[width=0.46\textwidth]{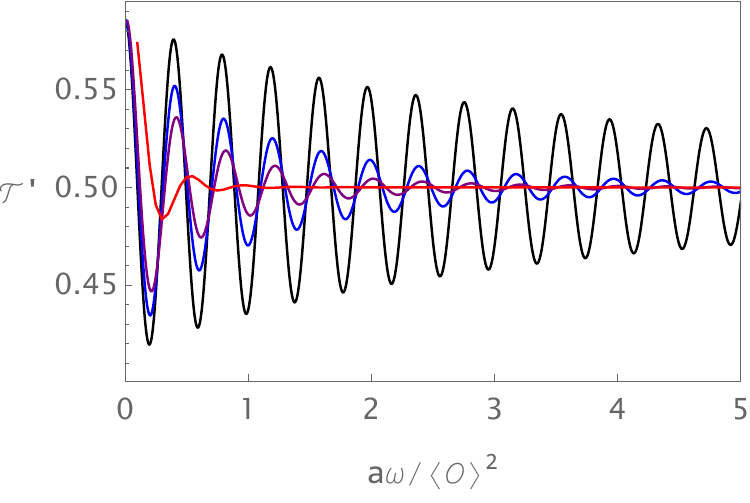}
~~~~~
\includegraphics[width=0.46\textwidth]{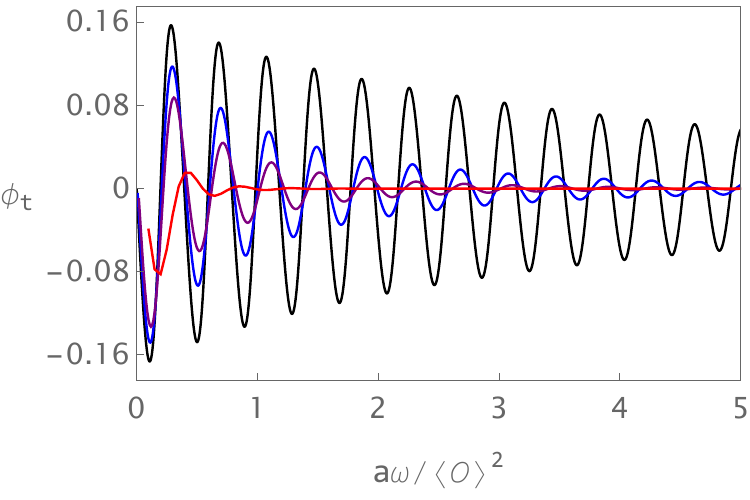}
\end{center}
\vspace{-0.3cm}
\caption{\small  The plots of  $\mathcal{T}'$ and $\phi_t$ as functions of $a \omega/\langle \mathcal{O}\rangle^2$, for different values of $b$ with fixed  $\omega/\langle \mathcal{O}\rangle^2=0.01$ ({\em upper two plots}) 
and for different values of $\omega/\langle \mathcal{O}\rangle^2$ with fixed $b=1$ 
({\em lower two plots}). 
In the 
lower two plots, 
$\omega/\langle \mathcal{O}\rangle^2=0.001$ ({\em black}), $0.005$ ({\em blue}),  $0.01$ ({\em purple}) and $0.05$ ({\em red}).
}
\label{fig:Tao}
\end{figure}

In Fig. \ref{fig:Tao}, we show the plots of $\mathcal{T}'$ and $\phi_t$ as functions of $a$. In the upper two plots, we fix the frequency and tune $b$. Both $\mathcal{T}'$ and $\phi_t$ exhibit oscillations with a period of approximately $\omega a/\langle \mathcal{O}\rangle^2\simeq 0.4$. 
{\color{black}
The amplitude of these oscillations decreases as $a\omega/\langle \mathcal{O}\rangle^2$  increases for fixed $b$. Furthermore, the 
upper two plots in Fig. \ref{fig:Tao} show that  
the oscillation amplitude also decreases when $b$ is increased at a fixed value of  $a\omega/\langle \mathcal{O}\rangle^2$.
}

\begin{figure}[h!]
\begin{center}
\includegraphics[width=0.46\textwidth]{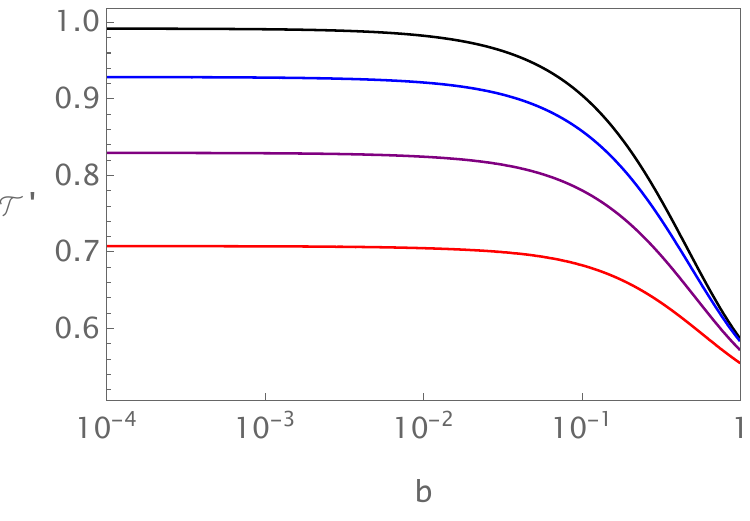}
~~~~~
\includegraphics[width=0.46\textwidth]{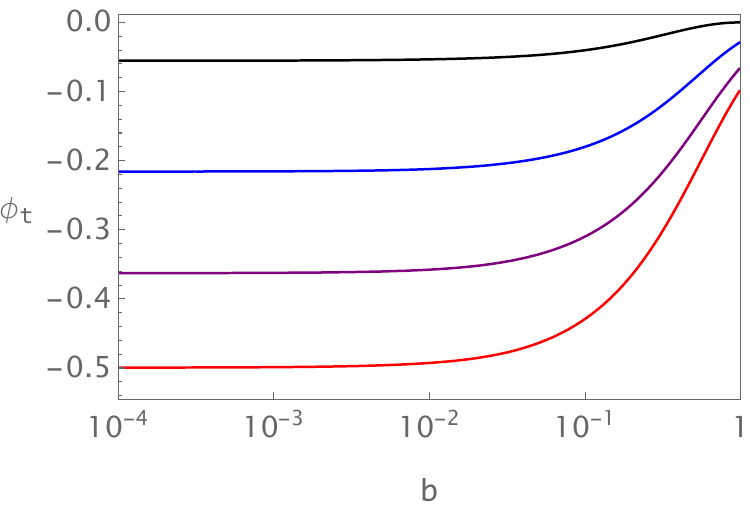}
\\
\includegraphics[width=0.46\textwidth]{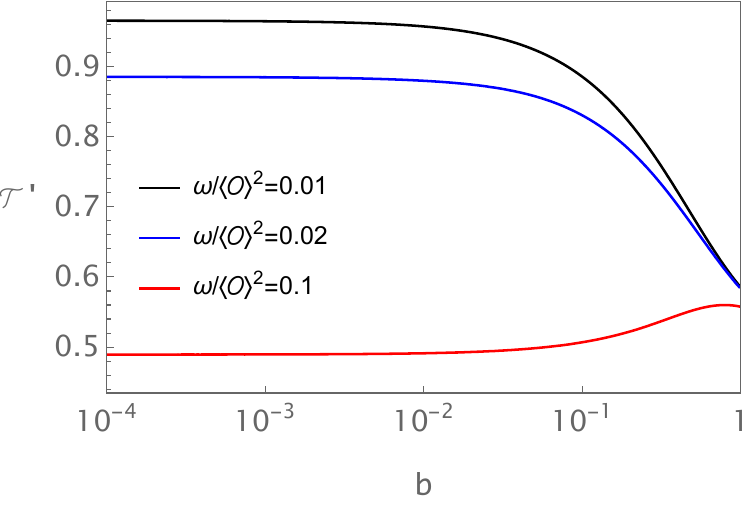}
~~~~~
\includegraphics[width=0.46\textwidth]{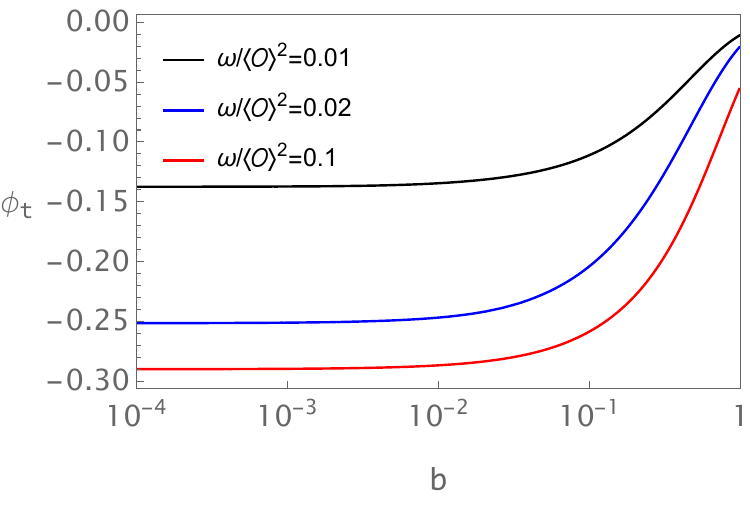}
\end{center}
\vspace{-0.3cm}
\caption{\small  The plots of  $\mathcal{T}'$ and $\phi_t$ as functions of $b$, for different values of $a$ with fixed $\omega/\langle \mathcal{O}\rangle^2=0.01$ ({\em upper two plots}) 
and for different values of $\omega/\langle \mathcal{O}\rangle^2$ with fixed $a=2$ 
({\em lower two plots}). 
In the upper two plots, $a=1$ ({\em black}), $3$ ({\em blue}), $5$ ({\em purple}), and $7$ ({\em red}).
} 
\label{fig:Tbo}
\end{figure}

In Fig. \ref{fig:Tbo}, we show the plots of $\mathcal{T}'$ and $\phi_t$ as functions of $b$. Due to the null energy condition  constraint $a\geq b$, we focus on the region where $b\in (0,1]$ with $a\geq 1$.  We observe that, for the parameters considered, $\mathcal{T}'$ and $\phi_t$ exhibit monotonic behavior for all  values of $b$ within this region. This behavior contrasts with the trends observed  in Fig. \ref{fig:Tao}. Moreover, we find that when $b\to 0$, $\mathcal{T}'$ approaches distinct values, which differs from its  dependence on $a$. 
Furthermore, for a fixed value of $b$, we observe that $\mathcal{T}'$ increases when $\omega/\langle \mathcal{O}\rangle^2$ decreases within the studied parameter regions,  and this is consistent with the observations in Fig. \ref{fig:Toab}  for the frequencies below the ``dip-ramp'' feature shown
there.\footnote{
We have  numerically verified that with a fixed   $a=2$ and various $b$ values in the range  $(0,1]$, the  transmission coefficient  $\mathcal{T}'$   decreases monotonically as $\omega/\langle \mathcal{O}\rangle^2$ increases across the interval  $[0.01,0.1]$.}

\subsection{An analytical example with UV and IR asymptotically AdS$_2$}

{\color{black}
In this subsection we show a brane profile for which analytical solutions of the transmission coefficients can be derived. In this example, the induced metric on the interface  brane is asymptotically AdS$_2$ in both the UV and IR regions. From the analytical solution, we observe key features that were previously identified in the numerical example in Sec.  \ref{subsec:sol1}. Another similar example is provided in the appendix \ref{app:c}.
}  


We consider the brane profile
\be
\psi(z)=\frac{a}{b}\log (1+b z) \, .
\ee 
The null energy condition requires that $ab> 0$. Here we consider $a>0$ and $b>0$. The background brane geometry is asymptotically AdS with radius $L\sqrt{1+a^2}$ in the UV 
and radius $L$ in the IR. 
From the UV expansion, it exactly matches with \eqref{eq:pzAB} by identifying  $A=a$ and $B=-ab/2$.

The profile for the scalar field is given by 
\be
\phi=\int_0^z ds \sqrt{
\frac{abL}{s(1+bs)\sqrt{a^2+(1+bs)^2}}
}    \,  ,
\ee
and the potential is 
\be
V(\phi(z))=\frac{
a(2+2a^2+5bz+3b^2z^2)
}{
L(a^2+(1+bz)^2)^{3/2}
}   \, .
\ee
In principle, this potential can be expressed as a function of $\phi$. The source of the scalar operator vanishes, and the VEV is $\langle \mathcal{O}\rangle=2\sqrt{abL}/(1+a^2)^{1/4}$. We have $V_\text{IR}\to 0$ and $V_\text{UV}\to 2a/(L\sqrt{1+a^2})$. 

The solution to the homogeneous part of ODE \eqref{eq:xiODE} takes the form  
\begin{align}
\label{eq:ana-log}
\begin{split}
\xi_l(z)=&\,~c_{-}\,
z e^{-i\sqrt{2}\omega z} (bz+1)^{\frac{ia\omega}{b}}\,
U\left(  
\frac{1}{2}-\frac{i(\sqrt{2}-4a)\omega}{4b},~
1+\frac{2ia\omega}{b},~
\frac{2i\sqrt{2}(1+bz)\omega}{b}
\right)     \,  \\
&+c_{+}\,
z e^{-i\sqrt{2}\omega z} (bz+1)^{\frac{ia\omega}{b}}
L\,\left(
-\frac{1}{2}+\frac{i(\sqrt{2}-4a)\omega}{4b},~
\frac{2ia\omega}{b},~
\frac{2i\sqrt{2}(1+bz)\omega}{b}
\right)   \,  
,
\end{split}
\end{align}
where $U(a,b,z)$ is the Tricomi confluent hypergeometric function, $L(a,b,z)$ is the generalized Laguerre polynomial and $c_{\pm}$ are constants. 
The infalling boundary condition at IR requires $ c_{+}=0  \, .$

From the solution \eqref{eq:ana-log} 
with $c_+=0$, 
matching the behavior of \eqref{eq:asyxil}, we obtain the transmission coefficient 
\begin{align} 
\mathcal{T}=1-
\frac{ab}{C}\,
U\left(
\frac{1}{2}-\frac{i(\sqrt{2}-4a)\omega}{4b},~
1+\frac{2ia\omega}{b},~
\frac{2i\sqrt{2}\omega}{b}
\right)   \,  ,
\end{align}
where
\begin{align}
\begin{split}
C=&\,~\sqrt{2}b\,
U\left(
\frac{1}{2}-\frac{i(\sqrt{2}-4a)\omega}{4b},~
1+\frac{2ia\omega}{b},~
\frac{2i\sqrt{2}\omega}{b}
\right)     \\
&+\left( \sqrt{2}b+(2\sqrt{2}a-1)i\omega \right)
U\left(
\frac{3}{2}-\frac{i(\sqrt{2}-4a)\omega}{4b},~
2+\frac{2ia\omega}{b},~
\frac{2i\sqrt{2}\omega}{b}
\right)   \,  
\end{split}
\end{align}
and \be
\mathcal{R}=1-\mathcal{T}\,,~~~~~~
c_{-}= \frac{2ab}{\omega^2 C}  \, .
\ee
Here we derive an analytic expression for the transport coefficients, which have a non-zero imaginary part and depend on $\omega$. Notably, $\mathcal{T}$ is determined by two parameters, $a$ and $\omega/b$. This is related to the fact that the IR behavior of the profile is special.  

As $a\to 0$, the transmission coefficient approaches $\mathcal{T}=1$, corresponding to a brane straightly perpendicular to the boundary. When $a\to +\infty$, we find $\mathcal{T}=1/2$ for different values of $\omega/b$ by numerical computation.  When $\omega/b\to 0$, due to the fact that 
\begin{align}
U\left( \frac{1}{2},\ 1,\ 2i\sqrt{2}\frac{\omega}{b} \right)
&\to -\frac{\gamma+\log\frac{i\omega}{\sqrt{2}b}}{\sqrt{\pi}}   \,  ,~~~~
\frac{\omega}{b}U\left( \frac{3}{2},\ 2,\ 2i\sqrt{2}\frac{\omega}{b} \right)
\to -\frac{i}{\sqrt{2\pi}}  \,  ,
\end{align}
where $\gamma$ is the Euler constant, 
and we have 
 $\mathcal{T}\simeq 1+\mathcal{O}(\frac{\omega}{b}\log \frac{\omega}{b})\to 1$, consistent with $V_\text{IR}=0$. When $\omega/b \to \infty$, numerical results indicate $\mathcal{T}\to 1+a^2-a\sqrt{1+a^2}$. 
Additionally, $\mathcal{T}$ exhibit oscillatory behavior with $\omega$, and for certain value of $a$ and $\omega/b$, the real part of $\mathcal{T}$ can fall below the lower bound of $1/2$ proposed in \cite{Bachas:2020yxv}. These features align with those observed in the example discussed  in Sec. \ref{subsec:sol1}.



\subsection{Another kind of UV asymptotically AdS geometry}
\label{subsec:IRflat}

From \cite{Liu:2024oxg} it is known that when the brane is asymptotic AdS$_2$ in the UV limit, the profile $\psi(z)$ must satisfy $\psi(z)\to a z^n$ with $n\geq 1$ as $z\to 0$. An interesting brane configuration is 
\be
\psi(z)=-a z^2  \, ,
\ee 
where the null energy condition requires that $a>0$. 
The induced metric of the brane is asymptotically AdS$_2$ in the UV and flat in the IR limit.

As noted in footnote  \ref{footnote:sourceless}, the source term in \eqref{eq:xiODE} vanishes. The exact solution to Eq. \eqref{eq:xiODE} is
\be
\xi(z)=c_{-}\,ze^{-i\omega a z^2}+c_{+}\,ze^{+i\omega a z^2}  \,  ,
\ee
where $c_{\pm}$ are two integral constants. The infalling boundary condition at IR requires 
\be
c_{+}=0 \,  .
\ee
Substituting the solution $\xi(z)=c_{-}ze^{-i\omega a z^2}$ into equation \eqref{eq:lambdaODE}, we find
\be
\lambda(0)=-\frac{2i\,c_-}{\omega} \,  .
\ee
The sourceless condition $\lambda(0)=0$ implies $c_{-}=0$, leading to 
\be
\xi(z)=0 \,  ,
\ee
which indicates the perturbation in the $u_\text{I,II}$ directions vanishes.

From Eq. \eqref{eq:psiODE}, the asymptotic behavior of $p(z)$ in the UV limit is
\begin{align}
p(z)=p_1\,\sqrt{z}\log(z)\left(1+\mathcal{O}(z)\right)
+p_2\,\sqrt{z}\left(1+\mathcal{O}(z)\right)
\,  ,
\end{align}
where $p_1$ and $p_2$ are two integration constants. The sourceless condition for $p(z)$ requires $p_1=0$.
Substituting $\xi(z)=0$ and the UV expansion of $p(z)$ with sourceless condition into Eqs.  \eqref{eq:h1ODE} and \eqref{eq:h2ODE}, we obtain
\begin{align}
h_\text{I}(0)=&-h_\text{II}(0)
+\frac{1+\mathcal{R}-\mathcal{T}}{\omega^2a}  \,  ,~~~~
h_\text{II}(0)=\frac{
2ia(\mathcal{R}+\mathcal{T}-1)
+(1+\mathcal{R}-\mathcal{T})\,\omega
}{
2a\omega^2
}  \,  .
\end{align}
Applying the sourceless conditions  $h_\text{I}(0)=h_\text{II}(0)=0$, the transport coefficients are found to be 
\be
\mathcal{R}=0\, ,\ \ \mathcal{T}=1 \,  .
\ee
Unlike the previous examples, the transmission coefficient is real and independent of the frequency of the injected waves. This result is identical to that of a topological interface, where the system is fully transmissive. Another similar result is obtained when the tension of the conformal brane vanishes \cite{Bachas:2020yxv}. It would be interesting to further explore other physical properties to understand the properties of this specific model. 
\vspace{0.25cm}

More generally, as discussed in Section \ref{subsec:bc}, one can consider a generic 
brane profile $\psi(z)$ with a UV expansion 
\be
\psi(z)=Az^2+Bz^3+\cdots  \, ,
\ee
where $A<0$. The solution to ODE  \eqref{eq:xiODE} has a UV expansion of the form 
\be
\xi(z)=\xi_1 z+\xi_3 z^3+
\frac{
B(1+\mathcal{R}-\mathcal{T}+\omega^2\xi_1+2\xi_3)
}{2A}z^4+\mathcal{O}(z^5)
\, .
\ee
The scalar perturbation $p(z)$, dual to a scalar operator in ICFT, has the UV expansion
\begin{align}
p(z)=p_1\sqrt{z}\log(z)\left(1+\mathcal{O}(z)\right)
+p_2\sqrt{z}\left(1+\mathcal{O}(z)\right)  
\,  ,
\end{align}
where $p_1$ and $p_2$ are integration  constants. Applying the sourceless condition $p_1=0$, the UV expansion of $p(z)$ simplifies to 
\be
p(z)=
p_2\sqrt{z}\left(1+\mathcal{O}(z)\right)
\, .
\ee

We first substitute $\psi(z),\xi(z)$ into Eq.  \eqref{eq:lambdaODE}, and find
\be
\lambda(0)= -\frac{2i\xi_1}{\omega }  \, .
\ee
The condition $\lambda(0)=0\, 
$
requires $\xi_1=0$. Second, substituting  $\psi(z),\xi(z)$ into \eqref{eq:h2ODE}, we obtain 
\be
h_\text{II}(0)=-h_\text{I}(0)-
\frac{
1+\mathcal{R}-\mathcal{T}+2\xi_3
}{A\omega^2}  \, .
\ee
The sourceless conditions 
$
h_\text{II}(0)=h_\text{I}(0)=0  \, 
$
imply  
\be
1+\mathcal{R}-\mathcal{T}+2\xi_3=0  \label{h2=0} \, .
\ee
Third, substituting $\psi(z),\xi(z), p(z)$ into \eqref{eq:h1ODE}, we find
\be
h_\text{I}(0)=
\frac{
2iA\,(\mathcal{R}+\mathcal{T}-1)-
(1+\mathcal{R}-\mathcal{T}+2\xi_3)\omega
}{
2A\omega^3
}
\, .
\ee
The condition  
$
h_\text{I}(0)=0  \, 
$
requires 
\be
2iA\,(\mathcal{R}+\mathcal{T}-1)-
(1+\mathcal{R}-\mathcal{T}+2\xi_3)\,\omega =0 \label{h1=0} \, .
\ee

Solving equations \eqref{h1=0} and \eqref{h2=0}, we obtain
\be
\mathcal{R}=-\xi_3  \, ,\ \ 
\mathcal{T}=1+\xi_3  \,  ,
\ee
which satisfy the relation 
\be
\mathcal{R}+\mathcal{T}=1 \,  .
\ee
Unitarity of the dual system constraints  $\xi_3\in [-1, 0]$. For the specific case $\psi(z)=-az^2$, the exact solution for $\xi(z)$ exists. The infalling boundary condition and $\xi_1=0$ imply $\xi_3=0$, leading to  $\mathcal{R}=0,\mathcal{T}=1$.


\section{Conclusions and open questions}
\label{sec:cd}

We have studied the properties of energy transport in a $1+1$-dimensional ICFT system with a non-conformal interface using holographic duality. Our focus has been on systems where the central charges of the left and right CFTs are identical, and the interface hosts a localized scalar operator. In the holographic framework, the system is described by a left and right bulk gravity joined along a brane anchored at the interface, with a dynamical scalar field localized on the brane. 

By imposing sourceless boundary conditions for the fluctuations on the interface brane, or equivalently enforcing the energy conservation at the interface in the dual theory, we consistently observe that the sum of transmission and reflection coefficients satisfies $\mathcal{R}+\mathcal{T}=1$.  Furthermore, the transmission coefficient $\mathcal{T}$ is generally complex. We interpret the real parts of these coefficients as the averaged transmission and reflection coefficients, and analyze their behavior across several brane profile configurations. Through one numerical and two analytical examples, we demonstrate that, in addition to the complex nature of the coefficients, the energy transports depend on the frequency of the injected wave, in contrast to  conformal interfaces, where these energy transport coefficients remain real constants independent of frequency. Notably, for cases where the induced metric on the interface brane interpolates between two AdS$_2$ geometries, the averaged transmission coefficient transitions from the UV conformal interface value at high frequencies to the  IR conformal interface value at low frequencies. In the intermediate frequency regime, the coefficient may oscillate as a function of frequency and can violate the bound found in \cite{Bachas:2020yxv}.
Additionally, in Sec. \ref{subsec:IRflat} we find a nontrivial example exhibiting complete transmissivity, which shares similarities with the behavior of a topological interface. 

Several open questions remain for future  exploration. A natural extension of this work would be to generalize our study to cases where the left and right CFTs have different central charges. Such scenarios are expected to involve more complicated numerics. One could also employ the non-perturbative methods \cite{Bachas:2021tnp}, such as studying the system at finite temperature, to investigate the brane profile and energy fluxes in detail.  Another interesting direction is to explore the real-time evolution of the system. One could excite a  localized energy flux on the left CFT and observe the system's dynamics, thereby elucidating the role of the complex and frequency-dependent energy transport properties.  Finally, it would be valuable to study energy transport directly from the  field theory perspective. While non-conformal interfaces in field theory have been explored in works such as \cite{Kim:2025tvu}, the properties of energy transport in these systems warrant further investigation. 
We hope to address some of these questions in future work.

\vspace{.3cm}
\subsection*{Acknowledgments}
 We thank Shan-Ming Ruan and 
 Ya-Wen Sun for useful discussions. This work is supported by the National Natural Science Foundation of China grant No. 12375041. 

\vspace{.6 cm}
\appendix
\section{The useful equations}
\label{app:a}
In this appendix we show additional equations that were not included in the main text but are essential for the analysis. 

\subsection{Equations of the system}
\label{app:a1}
We begin by writing out the equations of motion for the fields in the system. These equations are fundamental for solving both  the background configuration and the fluctuations of the system. 

The continuous condition for the metric on $Q$ is given by 
\be
\label{eq:continuous}
g_{ab}\frac{\partial x_\text{I}^a}{\partial y^\mu}
\frac{\partial x_\text{I}^b}{\partial y^\nu}\bigg{|}_\text{I}=g_{ab}\frac{\partial x_\text{II}^a}{\partial y^\mu}
\frac{\partial x_\text{II}^b}{\partial y^\nu}\bigg{|}_\text{II}\,.
\ee
The equations of motion are
\bea
\label{eq:einNI}
R^{\text{I}}_{ab}-\frac{1}{2}g^{\text{I}}_{ab}R^{\text{I}}-\frac{1}{L^2}g^{\text{I}}_{ab}&=&0\,,\\
\label{eq:einNII}
R^\text{II}_{ab}-\frac{1}{2}g^\text{II}_{ab}R^\text{II}-\frac{1}{L^2}g^\text{II}_{ab}&=&0\,,\\
\Delta K_{\mu\nu}-h_{\mu\nu}\Delta K
+\left[  
(\partial\phi)^2+V(\phi)
\right]
h_{\mu\nu}-2\partial_{\mu}\phi\partial_{\nu}\phi&=&0\,, \label{eq:Q1} \\
2\partial_{\mu}(\sqrt{-h}h^{\mu\nu}\partial_{\nu}\phi)-\sqrt{-h}\frac{dV(\phi)}{d\phi}&=&0\,,\label{eq:Q2} 
\eea
where $\Delta X\equiv X_{\text{I}}-X_\text{II}$ with $X$ as $K_{\mu\nu}$ or $K$. Equations  \eqref{eq:einNI} and \eqref{eq:einNII} describe the metric fields in the left half and the right bulk regions, respectively,  while \eqref{eq:Q1} and \eqref{eq:Q2} are the equations on the interface brane $Q$. Sepcifically, \eqref{eq:continuous} and \eqref{eq:Q1} are the generalized Isreal junction conditions. 

\subsection{Equations for the fluctuations}
\label{app:eom}

{\color{black}
From Eqs. \eqref{eq:continuous}  \eqref{eq:Q1} and \eqref{eq:Q2}, seven equations for the fluctuations can be obtained. 
Among them, only five equations  are independent. More precisely, Eq. \eqref{eq:Q2} and the $zz$  component of \eqref{eq:Q1} can be derived from the other five equations. These five independent equations can be simplified as Eqs.  \eqref{eq:xiODE}, \eqref{eq:psiODE}, \eqref{eq:lambdaODE}, \eqref{eq:h1ODE} and \eqref{eq:h2ODE}.
}

In Sec. \ref{sec:eqpertur}, we have written out the equations of motion for $\xi(z)$ and $p(z)$. The coefficients $c_1, c_2, c_3$ in Eq. \eqref{eq:psiODE} are
\begin{align}
\begin{split}
c_1&=
\frac{1}{4} \left[
\frac{1}{z^2}+4 \omega ^2(1+\psi'^2)+
\frac{4 \psi' \psi''}{z (\psi'^2+1)}+
\frac{(2-5 \psi'^2) \psi''^2}{(\psi'^2+1)^2}+
\frac{6 \psi' \psi^{(3)} }{\psi'^2+1}+
\frac{(\psi^{(3)})^2}{\psi''^2}-
\frac{2(z \psi^{(4)}+\psi^{(3)})}{z \psi''}
\right]  \, ,\\
c_2&=-\frac{z \psi^{(3)} \psi' (\psi'^2+1)+(\psi'^2+1) \psi'' (\psi'-i \omega  z)+z(1-\psi'^2)\psi''^2}
{2 \omega  \sqrt{z} (\psi'^2+1)^{5/4} \sqrt{-L \psi''}}   \,  ,\\
c_3&=\frac{z \psi^{(3)} \psi' (\psi'^2+1)+(\psi'^2+1) \psi'' (\psi'+i \omega  z)+z(1-\psi'^2)\psi''^2}
{2 \omega  \sqrt{z} (\psi'^2+1)^{5/4} \sqrt{-L \psi''}}  \,  .
\end{split}
\end{align}

Below, we provide the detailed form of three equations of  $\lambda(z),h_\text{I}(z),h_\text{II}(z)$, 
\begin{align}
\begin{split}
\lambda=&\,~
\frac{(\bm{I}+\bm{R}-\bm{T})z^3+2\xi}{i\omega z}  \,  ,\\
h_\text{I}=&\,~
\frac{
(1-\omega^2 z^2)\xi-z \xi'
}
{\omega^2 \psi'z^2}
+\sqrt{
\frac{-Lz}{\psi''}
}
\frac{
(1+\psi'^2)\left(\psi''+z\psi^{(3)}\right)-z\psi'\psi''
}{
\omega^2 (1+\psi'^2)^{7/4}z
}p  \,  \\
&+\frac{
2\sqrt{-Lz\psi''}
}{
\omega^2 (1+\psi'^2)^{3/4}
} p'
+\frac{1}{\omega^3 \psi'(1+\psi'^2)}\cdot\text{\bf s}
\,  , \\
h_\text{II}=&\,-h_\text{I}
+\frac{2(1-\omega^2 z^2)\xi-2 z \xi'}{\omega^2 \psi'z^2}    
-\frac{z
\left[
2(\bm{I}+\bm{R}-\bm{T})+i(\bm{I}-\bm{R}+\bm{T})\omega \psi' z
\right]
}
{\omega^2 \psi'}   \,  ,
\end{split}
\end{align}
where
\begin{align}
\begin{split}
\text{\bf s}=&
\left[ (\bm{T}-\bm{R})\omega z-\bm{I}   \right]
+i\left[  \bm{T}+(\bm{R}-\bm{I})(1+\omega^2 z^2) \right]
\psi'(1+\psi'^2)     \\
&+(\bm{I}+\bm{R}+\bm{T})\omega \psi'^4 z
+\left[ 2\omega\bm{T}+i(\bm{R}+\bm{T}-\bm{I})\psi''  \right] \psi'^2 z    \, .
\end{split}
\end{align}

\subsection{Equations for the boundary conditions}
\label{app:a3}
In this appendix we show the IR  boundary condition for the scalar field in Sec. 
\ref{subsec:sol1}. 
From \eqref{eq:psiODE}, the
asymptotic behavior of $p(z)$ in IR limit is
\begin{align}
\begin{split}
p(z)=&\,~ p_0 e^{-i\omega\sqrt{1+b^2}z}\left(
1-\frac{
i(1-ab\omega^2+b^2\omega^2)
}{
\sqrt{1+b^2}\omega z
}
+\mathcal{O}(z^{-2})
\right)    \\
&+e^{+i\omega bz}\left(
\frac{
e^{i(a-b)\omega}\sqrt{a-b}\,(\mathcal{R}+\mathcal{T})
}{
\sqrt{2L}\omega^2 (1+b^2)^{1/4} z
}
+\mathcal{O}(z^{-2})
\right)  \\
&+e^{-i\omega bz}\left(
\frac{
e^{-i(a-b)\omega}\sqrt{a-b}
}{
\sqrt{2L}\omega^2 (1+b^2)^{1/4} z
}
+\mathcal{O}(z^{-2})
\right)   \,  ,
\end{split}
\end{align}
where $p_0$ is an integration constant which can be determined by the sourceless boundary condition. We have dropped the out-going wave $e^{+i\omega\sqrt{1+b^2}z}$.

\section{Briefly review of the transport in conformal interface} 
\label{app:b}
In this appendix, we show that the gauge choice considered in Sec. \ref{sec:eqpertur} allows us to consistently reproduce the results for energy transports in conformal interface \cite{Bachas:2020yxv}.

The bulk equation is solved by the planar AdS$_3$ solution
\begin{align}
ds^2_A=\frac{L^2_A}{u^2_A}\bigg[-dt^2_A+
dx^2_A+du^2_A
\bigg]\,, ~~~ A=\text{I\,,~~II}\,.
\end{align}
The interface brane is described by 
\begin{align}
x_\text{I}=u_\text{I}\tan\alpha, \ \ 
x_\text{II}=-u_\text{II}\tan\beta.
\end{align}
The junction condition gives
\begin{align}
\nu\equiv\frac{L_\text{II}}{L_\text{I}}
=\frac{\cos\beta}{\cos\alpha}
\, ,~~~~ \sigma=\frac{\sin (\alpha+\beta)\sec\beta}{L_\text{I}}\, ,
\end{align}
where $\sigma$ is the brane tension. 
These two equations determine the value of $\alpha$ and $\beta$.

Since the study in the main text focuses on the case $\nu=1$, we generalize the fluctuations to arbitrary $\nu$. 
The first order perturbation of the bulk metric is given by
\begin{align}
\begin{split}
[ds^2]_\text{I}=&4GL_\text{I}\epsilon
\left[
e^{i\omega (t_\text{I}-x_\text{I})}(dt_\text{I}-dx_\text{I})^2+
\mathcal{R}e^{i\omega (t_\text{I}+x_\text{I})}(dt_\text{I}+dx_\text{I})^2
\right]  \,  , \\
[ds^2]_\text{II}=&4GL_\text{I}\nu\,\epsilon
\mathcal{T}e^{i\omega (t_\text{II}-x_\text{II})}(dt_\text{II}-dx_\text{II})^2  \,  ,
\end{split}
\end{align}
where $\mathcal{R}$ and $\mathcal{T}$ are the  reflection and transmission coefficients, respectively, and $\epsilon$ is the injected  energy flux.
The perturbed brane is described by 
\begin{align}
\begin{split}
t_\text{I}=&\,~t+\frac{G\epsilon}{L_\text{I}}e^{i\omega\tau}\lambda(z)\, ,\ 
x_\text{I}=z\tan\alpha+
\frac{2G\epsilon}{L_\text{I}}e^{i\omega\tau}h_\text{I}(z)\, ,\ 
u_\text{I}=z+\frac{2G\epsilon}{L_\text{I}}e^{i\omega\tau}\xi(z)\,  ,\\
t_\text{II}=&\,~t-
\frac{G\epsilon}{L_\text{I}}e^{i\omega\tau}\lambda(z)\, ,\ 
x_\text{II}=-z\nu\tan\beta +
\frac{2G\epsilon}{L_\text{I}}e^{i\omega\tau}h_\text{II}(z)\, ,\ 
u_\text{II}=\nu z-\frac{2G\epsilon}{L_\text{I}}e^{i\omega\tau}\xi(z)\,   ,
\end{split}
\end{align}
where $t$ and $z$ are the intrinsic coordinates of the brane.

Defining 
\begin{align}
\delta(z)\equiv\sin\alpha\,\csc\beta\, h_\text{I}(z)
+h_\text{II}(z)  \, ,
\end{align}
the perturbed equation of motion reduce to a system of four first-order differential equations for fields $\lambda(z),\ \xi(z),\ h_\text{I}(z),\  \delta(z)$. 
To solve this system, 
we impose the sourceless boundary conditions at $z=0$ \cite{Bachas:2020yxv} 
\begin{align}
\lambda(0)=0,\ 
\xi(0)=0,\ 
h_\text{I}(0)=0,\ 
\delta(0)=0\,  ,
\end{align}
ensuring that the intersection point between the brane and the boundary remains fixed. Notably, $\xi(0)=0$ is automatically satisfied. Additionally, we impose ``no out-going wave" condition, which eliminates  out-going wave $e^{i\omega(t+\sec\alpha\, z)}$. 
With these boundary conditions, we can fix the four functions and the transports $\mathcal{R}, \mathcal{T}$ are uniquely determined. 
From the solution, we obtain the holographic energy transport coefficients 
\begin{align}
\mathcal{T}=\frac{
2\cos\beta
}{
\cos\beta (1+\sin\alpha)+
\cos\alpha (1+\sin\beta)
}  \,  ,~~~~
\mathcal{R}=1-\mathcal{T}\, . \label{BachasT}
\end{align}

For the case $L_\text{I}=L_\text{II}$, the transmission coefficient $\mathcal{T}$ lies in the interval $[\frac{1}{2},1]$. The brane configuration corresponding to  $\mathcal{T}=1$ is given by $x_\text{I}=x_\text{II}=0$, while the brane configuration for $\mathcal{T}=\frac{1}{2}$  coincides with the AdS boundary at $z=0$. 

\section{Another solution with UV and IR asymptotically AdS$_2$
} 
\label{app:c}

{\color{black}

We consider another brane profile 
\be
\psi(z)=\frac{2a}{b}(\sqrt{1+bz}-1) \, .
\ee 
The null energy condition requires $ab>0$ and we again assume $a>0$ and $b>0$. The background brane geometry is asymptotically AdS with radius $\sqrt{1+a^2} L$ in the UV and $L$ in the IR. From the UV expansion, it exactly matches \eqref{eq:pzAB} by identifying that $A=a,B=-ab/4$.

The profile for the scalar field is 
\be
\phi=\int_0^z ds\,\sqrt{
\frac{abL}{2s(1+bs)\sqrt{1+a^2+bs}}
}    \,  ,
\ee
and the potential is 
\be
V(\phi(z))=\frac{
a(4+4a^2+5bz)
}{
2L(1+a^2+bz)^{3/2}
}   \, .
\ee
The source of the scalar operator vanishes,  and the VEV is $\langle \mathcal{O}\rangle=\sqrt{2abL}/(1+a^2)^{1/4}$. We have $V_\text{IR}\to 0$ and $V_\text{UV}\to 2a/(L\sqrt{1+a^2})$. 

The solution to the homogeneous part of ODE  \eqref{eq:xiODE} takes the form 
\begin{align}
\label{eq:ana-ads-2}
\begin{split}
\xi_l(z)=&\,~
c_{-}e^{-i\sqrt{\frac{3}{2}}\omega z}z\sqrt{1+bz}
U\left(
\frac{3}{4}+\frac{i(2a^2-1)\omega}{2\sqrt{6}b},\,
\frac{3}{2},\,
\frac{i\sqrt{6}(1+bz)\omega}{b}
\right)
\\
&+c_{+}e^{-i\sqrt{\frac{3}{2}}\omega z}z\sqrt{1+bz}
L\left(
-\frac{3}{4}-\frac{i(2a^2-1)\omega}{2\sqrt{6}b},\,
\frac{1}{2},\,
\frac{i\sqrt{6}(1+bz)\omega}{b}
\right)    
\,  ,
\end{split}
\end{align}
where $U(a,b,z)$ is the Tricomi confluent hypergeometric function, $L(a,b,z)$ is the generalized Laguerre polynomial and $c_{\pm}$ are constants. 
The infalling boundary condition at the IR requires 
$
c_{+}=0  \, .
$

From the solution \eqref{eq:ana-ads-2} 
with $c_+=0$, we obtain the transmission coefficient 
\begin{align}
\mathcal{T}=1+\frac{4iab\omega}{C_2}
U\left(
\frac{3}{4}+\frac{i(2a^2-1)\omega}{2\sqrt{6}b},~
\frac{3}{2},~
\frac{i\sqrt{6}\omega}{b}
\right)    \,  ,
\end{align} 
where
\begin{align}
\begin{split}
C_2=&\,~ 2b\left[
b-i(2a+\sqrt{6})\omega
\right]
U\left(
\frac{3}{4}+\frac{i(2a^2-1)\omega}{2\sqrt{6}b},~
\frac{3}{2},~
\frac{i\sqrt{6}\omega}{b}
\right)       \\
&+(4a^2\omega-2\omega-3i\sqrt{6}b)\omega
\,U\left(
\frac{7}{4}+\frac{i(2a^2-1)\omega}{2\sqrt{6}b},~
\frac{5}{2},~
\frac{i\sqrt{6}\omega}{b}
\right)        \,  ,
\end{split}
\end{align}
and 
\begin{align}
\mathcal{R}=1-\mathcal{T}   \,  ,\  \
c_{-}=-\frac{8iab}{\omega C_2}  \,  .
\end{align}
Here, $\mathcal{T}$ is also determined by two parameters, $a$ and $\omega/b$, similar to the previous analytical example.  

When $a\to 0$, we find that $\mathcal{T}=1+\mathcal{O}(a)$. 
When $a\to +\infty$, numerical computation show $\mathcal{T}=1/2$ for different $b,\omega$. 
When $\frac{\omega}{b}\to 0$, we can also find $\mathcal{T}\to 1$ and this rustlt is consistent with $V_\text{IR}=0$. 
When $\frac{\omega}{b}\to\infty$, we have $\mathcal{T}\to 1+a^2-a\sqrt{1+a^2}$ from numerical calculations. Oscillation behavior is also found for $\mathcal{T}$ in the intermediate frequency region with some parameters.   
And for certain value of $a$ and $\omega/b$, the real part of $\mathcal{T}$ can fall below the lower bound of $1/2$ proposed in \cite{Bachas:2020yxv}
These properties again align with the key features discussed in earlier sections. 

}



\vspace{.5cm}
\begin{thebibliography}{99}


\bibitem{Affleck:1991tk}
I.~Affleck and A.~W.~W.~Ludwig,
{\em Universal noninteger `ground state degeneracy' in critical quantum systems,} 
\doi{10.1103/PhysRevLett.67.161}{Phys. Rev. Lett. \textbf{67} (1991), 161-164}.

\bibitem{Sakai:2008tt}
K.~Sakai and Y.~Satoh,
{\em Entanglement through conformal interfaces,} 
\doi{10.1088/1126-6708/2008/12/001}{JHEP \textbf{12} (2008), 001} 
[\arXiv{0809.4548}{hep-th}].

 
\bibitem{Quella:2006de}
T.~Quella, I.~Runkel and G.~M.~T.~Watts,
{\em Reflection and transmission for conformal defects,}
\doi{10.1088/1126-6708/2007/04/095}{JHEP \textbf{04} (2007), 095}
[\arXiv{hep-th/0611296}{hep-th}].

\bibitem{Meineri:2019ycm}
M.~Meineri, J.~Penedones and A.~Rousset,
{\em Colliders and conformal interfaces,}
\doi{10.1007/JHEP02(2020)138}{JHEP \textbf{02} (2020), 138}
[\arXiv{1904.10974}{hep-th}]. 


\bibitem{Karch:2024udk}
A.~Karch, Y.~Kusuki, H.~Ooguri, H.~Y.~Sun and M.~Wang,
{\em Universal Bound on Effective Central Charge and Its Saturation,} 
\doi{10.1103/PhysRevLett.133.091604}{Phys. Rev. Lett. \textbf{133} (2024) no.9, 091604} [\arXiv{2404.01515}{hep-th}]. 

\bibitem{Oshikawa:1996ww}
M.~Oshikawa and I.~Affleck,
{\em Defect lines in the Ising model and boundary states on orbifolds,} 
\doi{10.1103/PhysRevLett.77.2604}{Phys. Rev. Lett. \textbf{77} (1996), 2604-2607} 
[\arXiv{hep-th/9606177}{hep-th}].

\bibitem{Lerner2008}
I.V. Lerner, V. I.  Yudson  and I. V. Yurkevich, 
{\em Quantum Wire Hybridized With a Single-
Level Impurity,} 
\doi{10.1103/PhysRevLett.100.256805
}{Phys. Rev. Lett. 100, 256805 (2008)}
\arXiv{0711.4919}{cond-mat.str-el}.

\bibitem{Rylands:2016gsf}
C.~Rylands and N.~Andrei,
{\em Quantum impurity in a Luttinger liquid: Exact solution of the Kane-Fisher model,} 
\doi{10.1103/PhysRevB.94.115142}{Phys. Rev. B \textbf{94} (2016), 115142} 
[\arXiv{1606.07472}{cond-mat.str-el}].

\bibitem{Billo:2016cpy}
M.~Bill\`o, V.~Gon\c{c}alves, E.~Lauria and M.~Meineri,
{\em Defects in conformal field theory,}
\doi{10.1007/JHEP04(2016)091}{JHEP \textbf{04} (2016), 091}
[\arXiv{1601.02883}{hep-th}].

\bibitem{Zaanen:2015oix}
J.~Zaanen, Y.~W.~Sun, Y.~Liu and K.~Schalm,
{\em Holographic Duality in Condensed Matter Physics,} 
\doi{10.1017/CBO9781139942492}{Cambridge Univ. Press (2015)}.

\bibitem{Hartnoll:2016apf}
S.~A.~Hartnoll, A.~Lucas and S.~Sachdev,
{\em Holographic quantum matter,}
MIT Press. 






\bibitem{Bachas:2020yxv}
C.~Bachas, S.~Chapman, D.~Ge and G.~Policastro, 
{\em Energy Reflection and Transmission at 2D Holographic Interfaces,}
\doi{10.1103/PhysRevLett.125.231602}{{Phys. Rev. Lett. \textbf{125} (2020) no.23, 231602}}
[\arXiv{2006.11333}{hep-th}].

\bibitem{Bachas:2021tnp}
C.~Bachas, Z.~Chen and V.~Papadopoulos,
{\em Steady states of holographic interfaces,} 
\doi{10.1007/JHEP11(2021)095}{JHEP \textbf{11} (2021), 095} 
[\arXiv{2107.00965}{hep-th}].


\bibitem{Baig:2022cnb}
S.~A.~Baig and A.~Karch,
{\em Double brane holographic model dual to 2d ICFTs,}
\doi{10.1007/JHEP10(2022)022}{JHEP \textbf{10} (2022), 022}
[\arXiv{2206.01752}{hep-th}].

\bibitem{Bachas:2022etu}
C.~Bachas, S.~Baiguera, S.~Chapman, G.~Policastro and T.~Schwartzman,
{\em Energy Transport for Thick Holographic Branes,} 
\doi{10.1103/PhysRevLett.131.021601}{Phys. Rev. Lett. \textbf{131} (2023) no.2, 021601}
[\arXiv{2212.14058}{hep-th}].

\bibitem{Brunner:2015vva}
I.~Brunner and C.~Schmidt-Colinet,
{\em Reflection and transmission of conformal perturbation defects,} 
\doi{10.1088/1751-8113/49/19/195401}{J. Phys. A \textbf{49} (2016) no.19, 195401} 
[\arXiv{1508.04350}{hep-th}].


\bibitem{Baig:2023ahz}
S.~A.~Baig and S.~Shashi,
{\em Transport across interfaces in symmetric orbifolds,}
\doi{10.1007/JHEP10(2023)168}{JHEP \textbf{10} (2023), 168}
[\arXiv{2301.13198}{hep-th}].

\bibitem{Biswas:2024xut}
P.~Biswas, S.~Das and A.~Dinda,
{\em Moving interfaces and two-dimensional black holes,}
\doi{10.1007/JHEP05(2024)329}{JHEP \textbf{05} (2024), 329}
[\arXiv{2401.11451}{hep-th}].


\bibitem{Baig:2024hfc}
S.~A.~Baig, A.~Karch and M.~Wang,
{\em Transmission coefficient of super-Janus solution,} 
\doi{10.1007/JHEP10(2024)235}{JHEP \textbf{10} (2024), 235} 
[\arXiv{2408.00059}{hep-th}].


\bibitem{Gutperle:2024yiz}
M.~Gutperle and C.~Hultgreen-Mena,
{\em Janus and RG-interfaces in minimal 3d gauged supergravity,}
[\arXiv{2412.16749}{hep-th}].

\bibitem{Liu:2024oxg}
Y.~Liu, H.~D.~Lyu and C.~Y.~Wang,
{\em On AdS$_{3}$/ICFT$_{2}$ with a dynamical scalar field located on the brane,} 
\doi{10.1007/JHEP10(2024)001}{JHEP \textbf{10} (2024), 001}
[\arXiv{2403.20102}{hep-th}].


\bibitem{Kanda:2023zse}
H.~Kanda, M.~Sato, Y.~k.~Suzuki, T.~Takayanagi and Z.~Wei,
{\em AdS/BCFT with brane-localized scalar field,}
\doi{10.1007/JHEP03(2023)105}{JHEP \textbf{03} (2023), 105}
[\arXiv{2302.03895}{hep-th}].


\bibitem{Banerjee:2024sqq}
A.~Banerjee, A.~Mukhopadhyay and G.~Policastro,
{\em Nambu-Goto equation from three-dimensional gravity,} 
\doi{10.1007/JHEP09(2024)013}{JHEP \textbf{09} (2024), 013} 
[\arXiv{2404.02149}{hep-th}].

\bibitem{Erdmenger:2014xya}
J.~Erdmenger, M.~Flory and M.~N.~Newrzella,
{\em Bending branes for DCFT in two dimensions,} 
\doi{10.1007/JHEP01(2015)058}{JHEP \textbf{01} (2015), 058}
[\arXiv{1410.7811}{hep-th}].

\bibitem{Erdmenger:2015xpq}
J.~Erdmenger, M.~Flory, C.~Hoyos, M.~N.~Newrzella, A.~O'Bannon and J.~Wu,
{\em Holographic impurities and Kondo effect,}
\doi{10.1002/prop.201500079}{Fortsch. Phys. \textbf{64} (2016), 322-329}
[\arXiv{1511.09362}{hep-th}].

\bibitem{Kim:2025tvu}
S.~Kim, P.~Kraus and Z.~Sun,
{\em Codimension one defects in free scalar field theory,} 
[\arXiv{2502.19547}{hep-th}].


\end{thebibliography}
\end{document}